\documentclass[useAMS,usenatbib]{mn2e}
\usepackage{rotating}
\usepackage{journals}
\usepackage{graphicx, subfigure}
\usepackage{xspace}
\usepackage{amssymb}

\def\farcs{\hbox{$.\!\!^{\prime\prime}$}}
\def\degr{\hbox{$^\circ$}}
\def\arcmin{\hbox{$^\prime$}}

\def\ra{$\rmn{RA}$}
\def\dec{$\rmn{Dec.}$}

\def\sec{$\rmn{s}$ }
\def\ang{$\rmn{\AA}$\xspace}

\def\chan{{\it Chandra}\xspace}

\def\src{EXO~0748$-$676\xspace}

\newcommand\msun {M$_{\odot}$\xspace}

\newcommand{\ergsec}{$\rmn{erg\,s^{-1}}$}

\LARGE \normalsize \title[]{Optical spectroscopy of the quiescent counterpart to \src: a Black Widow scenario?}

\author[Ratti et al.]  {E.M.~Ratti$^{1}$\thanks{email : e.m.ratti@sron.nl},
 D.T.H.~Steeghs$^{2,3}$, P.G.~Jonker$^{1,2,4}$, M.A.P.~Torres$^{1,2}$, C.G.Bassa$^{5}$, \newauthor F. Verbunt$^{6,1}$ \\ 
$^1$SRON, Netherlands Institute for Space Research,
Sorbonnelaan 2, 3584~CA, Utrecht, The Netherlands\\ 
$^2H$arvard--Smithsonian Center for Astrophysics, 60 Garden Street, Cambridge, MA~02138, U.S.A.\\
$^3$Department of Physics, University of Warwick, Coventry CV4 7AL\\
$^4$Department of Astrophysics, IMAPP, Radboud University Nijmegen, Heyendaalseweg 135,6525 AJ, Nijmegen, The Netherlands \\
$^5$Jodrell Bank Centre for Astrophysics, The University of Manchester, Manchester M13 9PL\\
$^6$Astronomical Institute, Utrecht University, PO Box 80 000, 3508 TA Utrecht, The Netherlands \\
}
\begin{document}

\maketitle

\begin{abstract} \noindent  We present phase-resolved optical spectroscopy of the counterpart to the neutron star low mass X-ray binary \src, almost one year after it turned into quiescence. The spectra display prominent H$\beta$ and H$\gamma$ and weak Fe\,{\sc ii} lines in emission. An average of all the spectra (corrected for the orbital motion) also exhibits a very weak line from Mg {\sc i}. Tomographic reconstructions show that the accretion disc is not contributing to the optical line emission, which is instead dominated by the irradiated hemisphere of the companion star facing the neutron star. We could not detect absorption features from the mass donor star in the spectra. The emission lines appear broad, with an intrinsic FWHM of 255$\pm$22$~\rmn{km\,s^{-1}}$. Under the assumption that the width of the Fe\,{\sc ii} emission lines is dominated by rotational broadening, we obtain a lower limit on the compact object mass which is inconsistent with a NS accretor.  We discuss this incongruity and conclude that either the lines are blends of unresolved features (although this requires some fine tuning) or they are broadened by additional effects such as bulk gas motion in an outflow. The fact that the  Fe\,{\sc ii}  lines slightly lag in phase with respect to the companion star can be understood as outflowing gas consistent with a Black-Widow like scenario. Nevertheless, we can not rule out the possibility that blends of various emission lines cause the apparent phase lag of the Fe\,{\sc ii} emission lines as well as their large width. 

\end{abstract}


\section{Introduction} 

\src was discovered with the {\it European X-Ray Observatory Satellite} (EXOSAT, \citealt{pawhgi1985}) in 1985. Soon after the discovery, the detection of type I X-ray bursts from the source \citep{gohapa1986} marked it as a Galactic  low-mass X-ray binary (LMXB) where a neutron star (NS) is accreting matter from a low-mass companion star. Many LMXBs are known to be transient, alternating periods of  quiescence at a relatively low X-ray luminosity ($\sim$10$^{32}$ \ergsec in a 0.5-10 keV range) with month to year-long bright X-ray  outbursts (10$^{36}-$10$^{38}$ \ergsec).  \src has been continuously in outburst for the 24 years since its discovery: tens of years-long X-ray outbursts have been observed from a number of LMXBs, e.g. KS~1731-260 \citep{2001ApJ...560L.159W}, GRS~1915$+$105 and 4U~1755$-$338 (see \citealt{2006ARA&A..44...49R} for a review). During outbursts mass accretion proceeds at a high rate forming an extended accretion disc around the accreting compact object. X-rays are emitted from the innermost regions, whereas the optical originates further out in the disk. In LMXBs the disc dominates the optical flux during outbursts, outshining  the low-mass companion star. The latter can become visible during quiescence, when the disc is less bright. 

\noindent A few LMXBs are known where the NS is detected as a pulsar  \citep{2005AIPC..797...71C} but for the majority of them the companion star is the only viable source of information regarding the system dynamics. Optical spectroscopy of the companion star can be used to measure the orbital parameters and, under certain conditions (see below), the mass of the accreting compact object.  This measure is interesting in LMXBs since the accretion process can significantly increase the mass of a NS: the maximum mass that a NS can reach is one of the parameters that can distinguish among the existing models for the equation of state (EoS) of those object \citep{2001ApJ...550..426L}.  The determination of the NS EoS is one of the key goals in the study of NSs and will have strong implication for both astronomy and super-nuclear density matter physics.

\noindent The mass of a NS (or a BH) in a non-pulsating LMXB can be measured by solving the system mass function \begin{center}
\begin{equation}
f(M_D)  = {{M_X}^3\sin^3i\over (M_D+M_X)^2} = M_X{sin^3i\over (1+q)^2} = {P{K_D}^3\over2\pi G}
\end{equation}
%$f(M_D):\frac{M_{X}sin^{3}i}{(1+q)^{2}}=\frac{K_D^{3}P}{2\pi G}$
\end{center} 
where $G$ is the universal constant of gravity, $M_D$ and $M_X$ are the masses of the companion and NS, respectively, $P$ is the orbital period, $i$ the inclination of the orbital plane, $K_D$ the amplitude of the radial velocity curve of the companion and where we define $q\equiv M_D/M_X$. This equation is valid provided that $K_D$ represents the motion of the center of mass of the companion. Because $f(M_D)\leq M_X$, the mass function provides a lower limit to the mass of the neutron star \citep{2003astro.ph..8020C}. 
%$P$ is the binary orbital period, K$_2$ is the radial velocity semi-amplitude of the companion star, $i$ is the inclination of the system and $q$ is the ratio between the mass M$_{D}$ of the companion and the NS mass M$_{X}$.

\noindent  $P$ and K$_D$ can be inferred from the orbital Doppler shift of stellar absorption lines originating in the atmosphere of the companion star, which can be visible in the optical spectra during quiescence. 
$f(M_D)$ can be solved for M$_{X}$  in eclipsing quiescent LMXBs \citep{1995xrbi.nasa...58V}, where $q$ can be expressed as a function of K$_D$ and of the  projected rotational velocity of the companion star $v\sin i= {2\pi\over P}R_D \sin i$ \citep{1988ApJ...324..411W} measured from the broadening of the companion stellar absorption lines \citep{1992oasp.book.....G}. From $q$ and from the eclipse duration, $i$ can be obtained \footnote{$\sin^2i\cos^2(\pi\Delta\phi)=1-[{0.49q^{2/3}\over0.6q^{2/3}+ln(1+q^{1/3})}]^2$ where $\Delta\phi$ is the eclipse duration (see \citealt{1985MNRAS.213..129H}). This relation is valid under the assumption that the companion is Roche Lobe filling, which holds for LMXBs.}. 
%Any small uncertainty in $i$ will translate in a small uncertainty in the mass determination due to the $\sin i$ cubed term in the mass function. 
%  If $q$ and $i$ are also known, $f(M_D)$ can be solved for M$_{X}$ \citep{1995xrbi.nasa...58V}. LMXBs are Roche Lobe filling systems, a condition that allows to express $q$ as a function of K$_2$ and of the  projected rotational velocity of the companion star $v\sin i$ \citep{1988ApJ...324..411W}. The latter can also be measured using optical spectroscopy, from the broadening of the companion  stellar absorption lines \citep{1992oasp.book.....G}. Finally,  $i$ can be best  obtained in an LMXB if X-ray eclipses are present: the eclipse duration is a function of $i$ and $q$. Furthermore, any small uncertainty in $i$ will translate in a small uncertainty in the mass determination due to the $\sin^3$ term in the mass function. Quiescent eclipsing LMXBs offer therefore the best conditions for an accurate NS mass measurements.

\noindent The LMXB  \src shows X-ray eclipses \citep{pawhgi1986} and, after a more than 20 year-long outburst, it turned into quiescence in September 2008 (\citealt{2008ATel.1736....1W}, \citealt{2008ATel.1812....1W}, \citealt{Hynes:2009p78}, \citealt{2008ATel.1817....1T}).  The optical counterpart to the X-ray source was first found by \citet{pawhgi1985} and confirmed by the \chan localisation of \src at \ra$=07^{\rmn{h}}48^{\rmn{m}} 33\fs73$, \dec$=-67\degr 45\arcmin 07\farcs 9$ \citep{2008ATel.1817....1T}. A search on photographic plates showed that the source was not detected down to $\sim$23 magnitudes when the X-ray emission was off. Spectroscopic observations in the optical during the outburst have  been performed by \citet{Pearson:2006p241} and \citet{MunozDarias:2009p73}, attempting to get a lower limit on the NS mass from  He, C and N emission lines probably originating from the inner, heated face of the companion star. The first spectroscopic study of \src in quiescence was performed by \citet{2009MNRAS.399.2055B} two months after the end of the outburst. The spectra, acquired in a 5750-7310 \ang wavelength range, were dominated by strong H$\alpha$ and weaker He emission lines coming from the companion star facing the NS, an indication of irradiation. A weak contribution to the lines was due to optical radiation from a residual accretion disk. No absorption lines were detected, preventing a straightforward solution of the $f(M_D)$. Nevertheless, the authors could put a lower limit on K$_D$ and on the NS mass, $M_{NS}>1.27$\msun.

 \noindent \citet{2009MNRAS.399.2055B} determined a minimum temperature for the irradiated companion star of \src of $\sim$5000 $\rmn{K}$, the surface temperature of a G type star. We have performed phase-resolved spectroscopic observations of the optical counterpart to \src after almost one year of quiescence. We attempted to detect absorption lines from the heated companion star by collecting our spectra in the wavelength range (4222-5701 \ang) where a $G$ as well as a $K$ type star would show absorption lines. We here report on the results of this study.
 \begin{figure}
\includegraphics[width=9.4cm, angle=0]{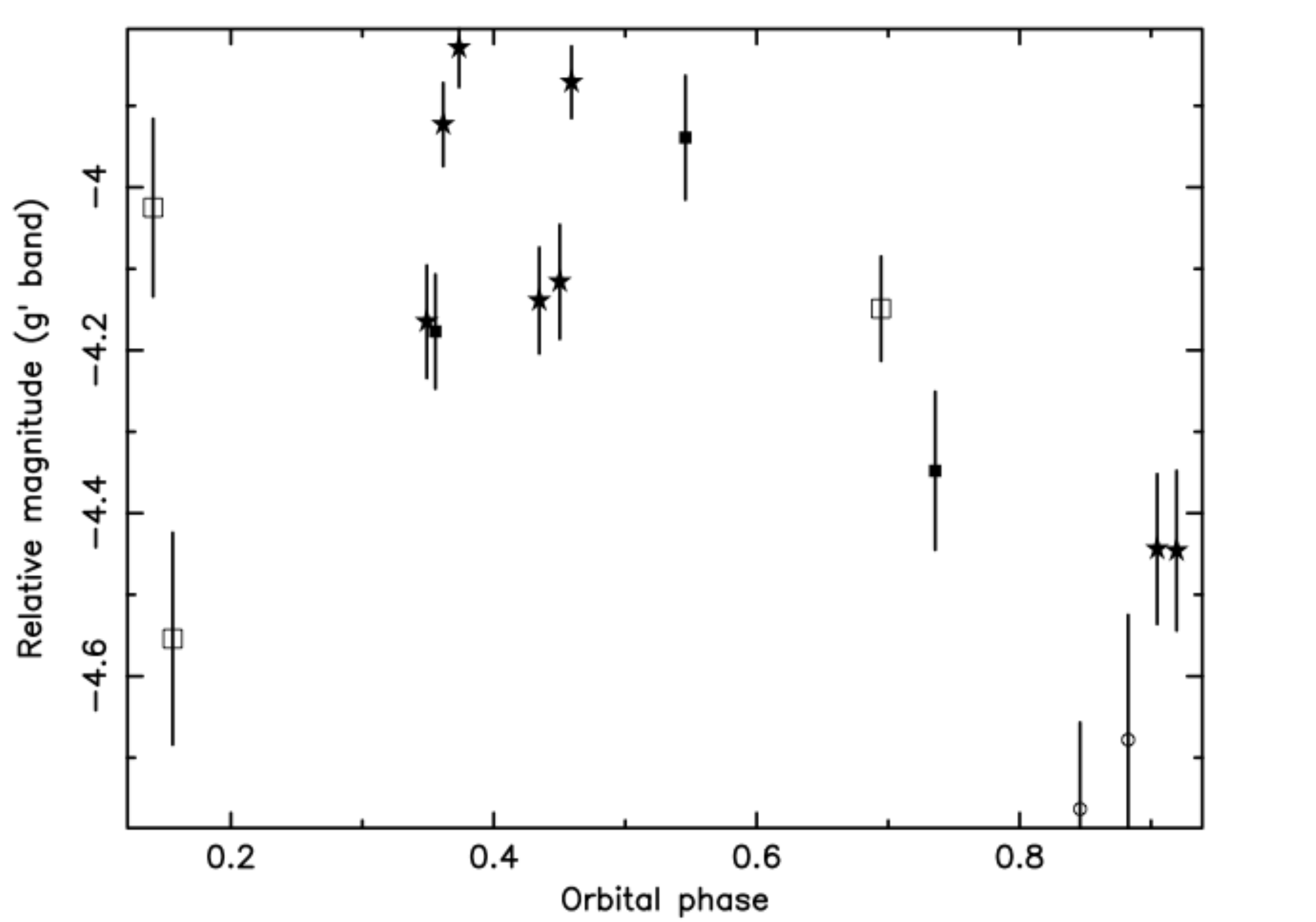}
\caption{g'-band light curve determined from the acquisition images. The magnitudes are relative to the bright star USNO-B1.0 0222$-$0189796. Different symbols refer to different observing nights starting from the night of 2010 January 18/19. In temporal order, the observations of the first to the fourth nights are indicated by black stars, black squares, empty circles and empty squares respectively.} 
\label{licu}
\end{figure}

\begin{figure*}
%\begin{tabular}{c}
\includegraphics[width=15cm, angle=0]{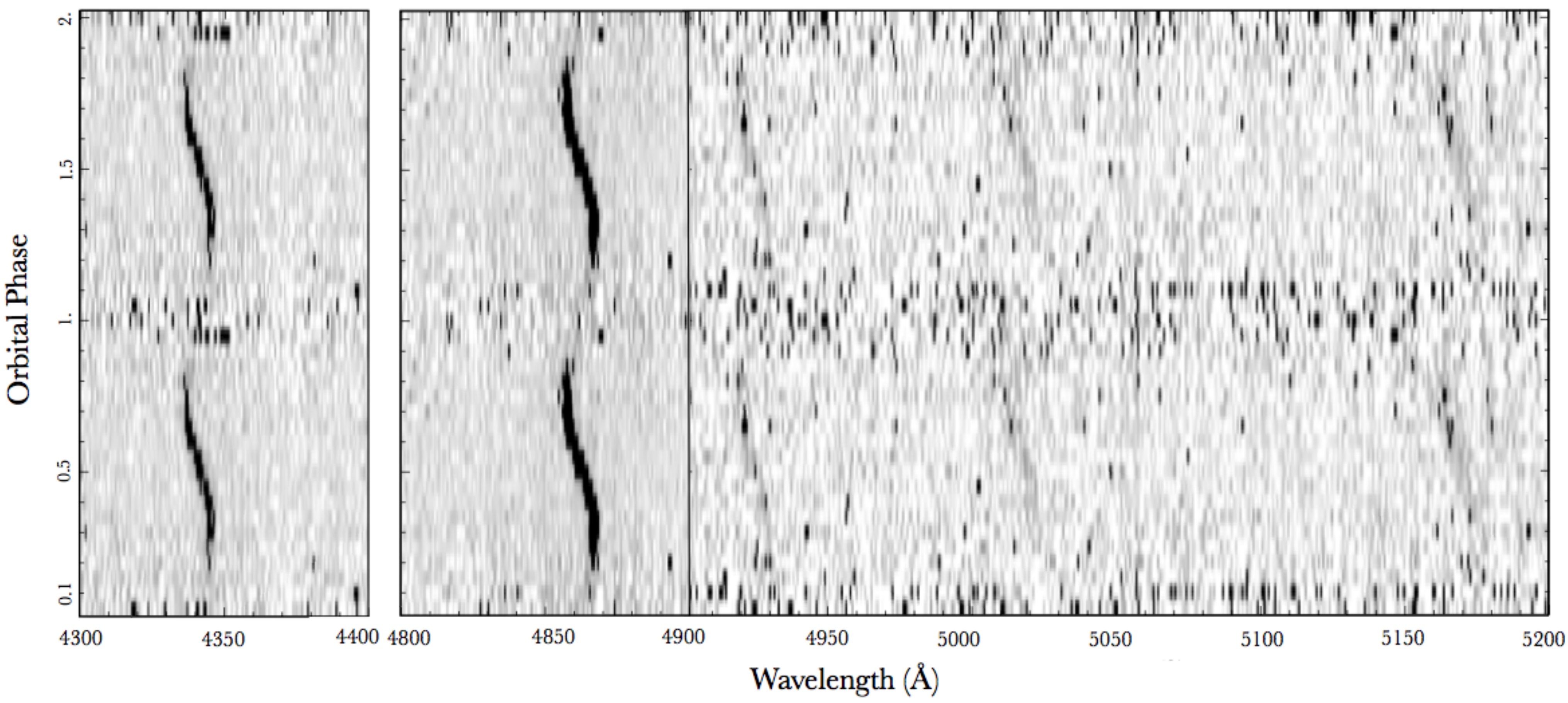} 
%\end{tabular}
\caption{ Trailed spectra composed by 73 single spectra of  the counterpart to \src, phase binned in 20 bins.  The two prominent s-waves are from the H$\gamma$ and H$\beta$ lines. Three weak lines consistent with Fe\,{\sc ii} are just visible between 4900 and 5200 \ang (see section \ref{features}).} 
\label{trail}
\end{figure*}

\section{Observations and data reduction}

We performed long-slit phase-resolved spectroscopy of the optical counterpart to \src with the FORS instrument on the Very Large Telescope (VLT) \footnote{VLT observing program 085.D-0441(C)} (grism 1200g+96 and a 1\farcs0 slit). The orbit was sampled with a total of seventythree 900\sec long exposures collected  between 2010 January 18 (MJD(UTC)$=$55214) and 2010 January 22, in the wavelength range 4222-5701 \ang. We also acquired template spectra from main sequence G type stars (G5V, G9V and G6V). 
The seeing across the four nights was between $\sim0\farcs8$ and $\sim1\farcs2$. The detectors were read out with a 2$\times$2 binning, providing a resolution of 2 \ang (measured from the width of both arc lines and of the night sky $O I$ line at 5577.338 \ang) sampled with a dispersion of 0.73 \ang $\rmn{pix^{-1}}$.   
The images were corrected for bias,  flat-fielded and extracted using the {\sc Figaro} package within the {\sc Starlink} software and  the packages {\sc Pamela} and {\sc Molly}  developed by T. Marsh.  We used  dome flats for the flat-fielding and we subtracted the sky continuum by fitting clean sky regions along the slit with a second order polynomial. The spectra were optimally extracted  following the algorithm of \citet{1986PASP...98..609H} implemented in  {\sc Pamela} and wavelength-calibrated in {\sc Molly} with a final accuracy of 0.1 \ang, using arc exposures  taken during daytime. The wavelength calibration was corrected for shifts in the single observation with respect to the position of the sky $O I$ line at 5577.338 \ang \citep{1996PASP..108..277O}.  Each spectrum has been normalised dividing by a first-order polynomial fit of the continuum. The spectra have been phase binned  with the $T_0=54776.501663\pm0.000068$ MJD/TDB from \citet{2009MNRAS.399.2055B}, which is the closest in time to our observations and with the orbital period $P_{orb}= 0.15933783446$ days from \citet{2009ApJS..183..156W}.  

\noindent We also analysed sixteen 6 s-long g'-band acquisition images, corrected for bias and flat-fielded with standard routines running in {\sc midas}. The photometry was performed through point spread function fitting, using {\sc daophot~II} \citep{1987PASP...99..191S}. Absolute photometric calibration was not possible due to the lack of observations of g'-band standard stars, but a light curve (Fig. \ref{licu}) of the optical counterpart to \src was obtained through relative photometry with respect to the reference star USNO-B1.0 0222$-$0189796. The instrumental magnitude of the reference star was measured with an accuracy of 0.03-0.08 magnitudes across the various images.  Relative photometry with respect to two other bright targets have shown that the reference star was not variable during our observations. The light curve was phase-folded with the same ephemeris used to phase-bin the spectra (see above).

\section{Analysis and results}

\subsection{Individual and trailed spectra}
\label{features}
The sole features detected in the individual spectra are the Balmer lines H$\beta$ (4861.327 \ang) and H$\gamma$ (4340.465 \ang), which are observed as strong emission features. We searched for  absorption lines in the individual spectra by cross-correlating with template star spectra, but we found no correlation. 

\noindent As shown in Figure \ref{trail}, more emission features become observable  when trailed spectra are constructed.  Besides the s-waves associated with the Balmer lines, three weak s-waves are visible in the region between 4900-5300 \ang : the position of the first two lines is consistent with a couple of He\,{\sc i} lines, at 4921.929 \ang and 5015.675 \ang, but also with Fe lI lines at 4923.92 \ang and 5018.44 \ang . No He\,{\sc i} line matches the position of the third line, which instead can be Fe\,{\sc ii} at 5169.03 \ang \citep{1972mtai.book.....M}. This favors an interpretation of the three lines as an Fe triplet as all the three lines are part of multiplet 42. 

\noindent The variations show only one peak along the orbit. The emission lines are fading away and disappearing between phase $\sim$0.75 and $\sim$0.25, consistent with an emission region located on the inner face of the companion star.  There is no indication of absorption lines in the trailed spectra.

\begin{figure*}
\centering
\begin{tabular}{cc}
\includegraphics[width=8cm, angle=0]{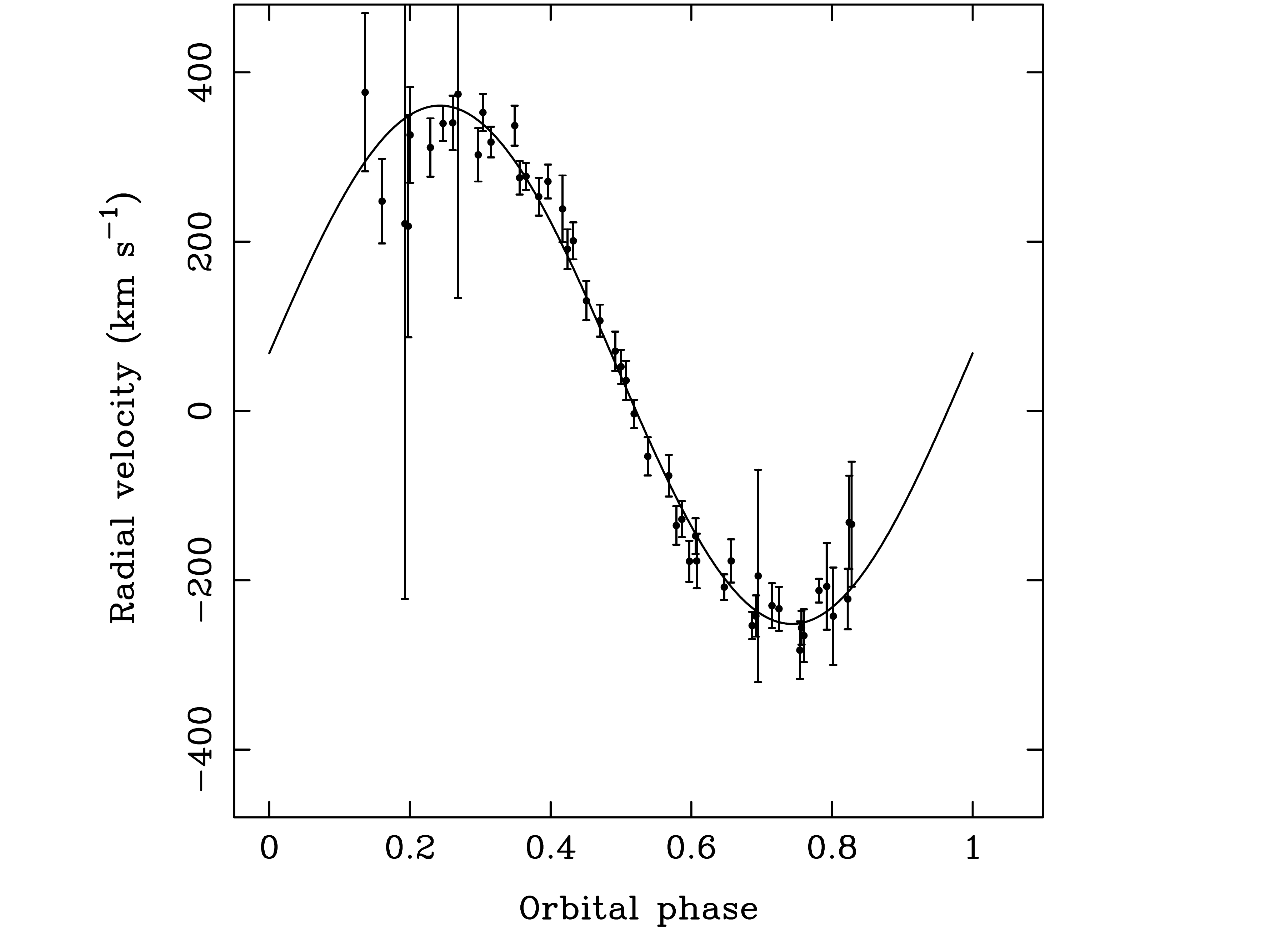} &
\includegraphics[width=8cm, angle=0]{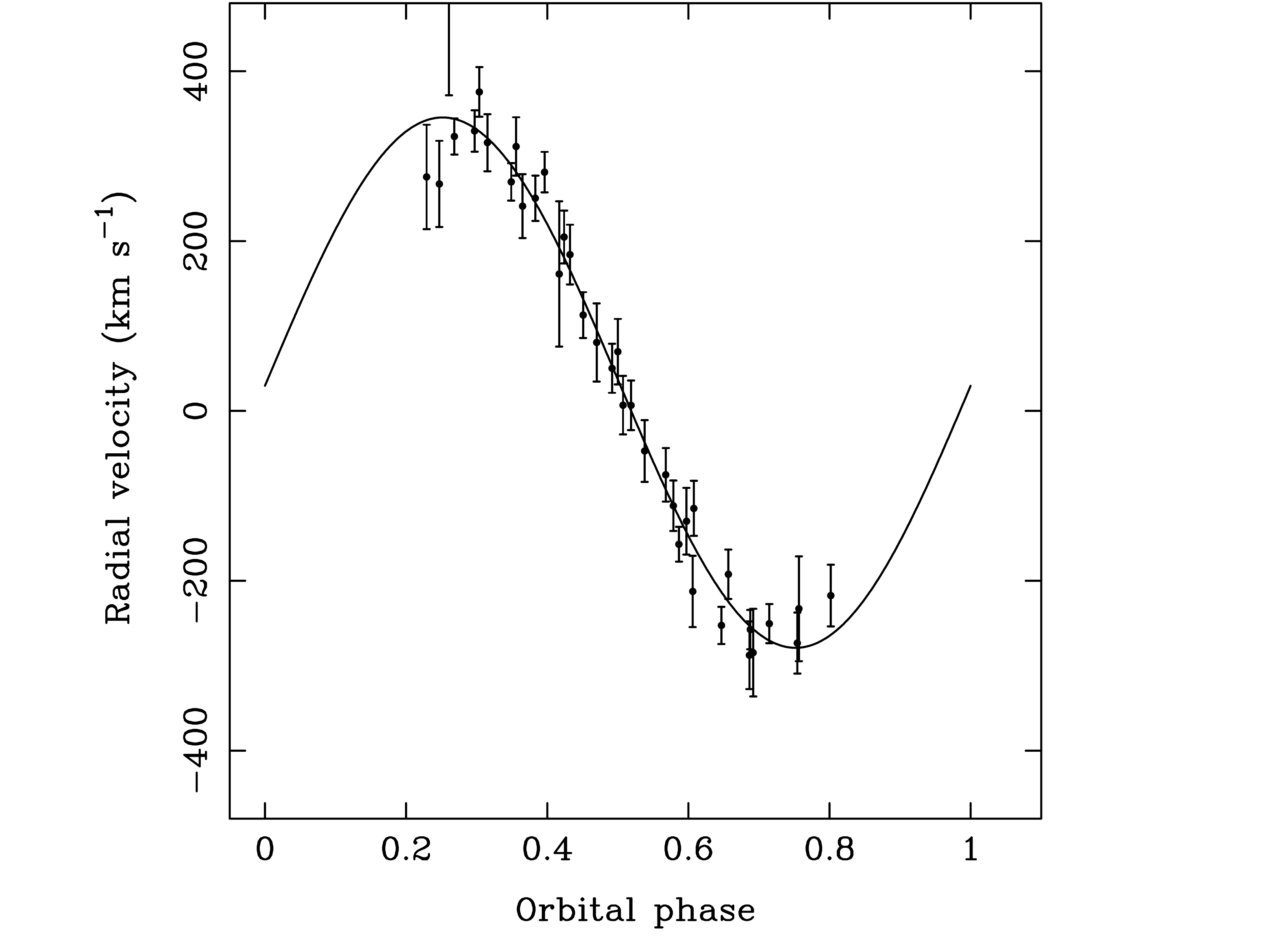} \\
(a) & (b) \\
\end{tabular}
\caption{ Radial velocity curves of the H$\beta$ (a) and H$\gamma$ (b) emission lines. The data points are obtained by fitting a Gaussian to each Balmer line in individual 900 s long spectra and selecting significant line detections at a 3$\sigma$ level. The solid lines show the best fit to the data.}
\label{rvc}
\end{figure*}

\subsection{Radial velocity of the Balmer lines}
\label{rvcsec}
For the H$\beta$ and H$\gamma$ lines we have obtained radial velocity curves (see Figure \ref{rvc}) by fitting single Gaussians to those lines in each 900 s spectrum,  including only  significant detections on a 3$\sigma$ level.
The fitted parameters were normalisation, full-width-at-half-maximum (FWHM) and velocity offset of the centroid with respect to the rest frame line wavelength. 
The 0.1 \ang 1$\sigma$ uncertainty on the velocity offsets due to the wavelength calibration  has been added in quadrature to the error on the centroid from the Gaussian fit of each observation.  We fitted the radial velocity curve  (velocity offset versus phase) for each line with a circular orbit in the form v($ \phi$)=$\gamma+$K$_\mathrm{em}sin(2\pi\phi +\varphi) $.  
We measured $\gamma=54.5\pm5.9~\rmn{km\,s^{-1}}$, K$_\mathrm{em}=306.1\pm5.0~\rmn{km\,s^{-1}}$ and $\varphi=0.007\pm0.004$ from the H$\beta$, $\gamma=32.8\pm9.7~\rmn{km\,s^{-1}}$, K$_\mathrm{em}=312.6\pm6.4~\rmn{km\,s^{-1}}$ and $\varphi=0.001\pm0.007$ from the H$\gamma$ line. The errors are obtained after we artificially increased the error-bars on the individual measurements such that the fit reduced $\chi^2$ was 1. The initial reduced $\chi^2$ was 1.6 for the H$\beta$ line (45 d.o.f.) and 2.2 for the H$\gamma$ one (33 d.o.f).
The fit  of the radial velocity curve of the two Balmer lines provides values consistent within 1$\sigma$  for  $\varphi$ and K$_\mathrm{em}$, and within 2$\sigma$  for $\gamma$. The latter is expected to be the same for all the lines, being the radial velocity of the system center of mass. $\varphi$ and K$_\mathrm{em}$ instead are not a priori the same for different emission lines, as they could originate from different parts of the irradiated companion star, or in fact regions in the binary.

\begin{figure*}
%\begin{minipage}[b]{8.5 cm}
\centering
\includegraphics[width=13 cm, height= 10cm, angle=0]{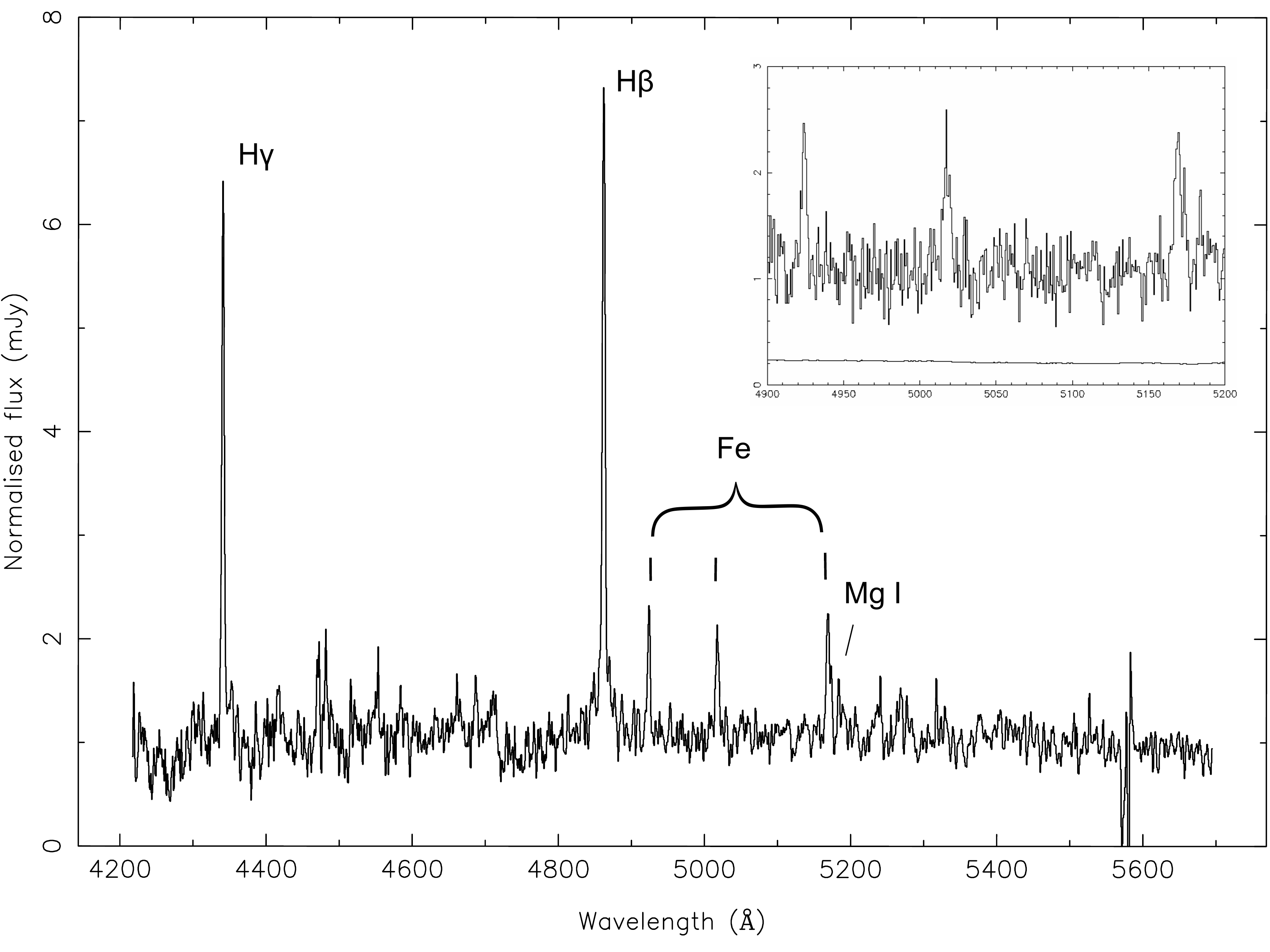}
\caption{ Spectrum of the counterpart to \src. This spectrum is an average of thirty-one 900 s-long spectra collected in the phase interval 0.3 and 0.7, corrected for the companion star orbital motion (see Section \ref{aversec}). The feature at $\sim$5577 \ang is due to non perfect subtraction of the night sky $O I$ line at that wavelength. The two prominent emission lines are from H$\gamma$ (4340.465 \ang) and H$\beta$ (4861.327 \ang). Four weaker lines are marked in the spectrum: the position of the first two is consistent with He\,{\sc i} lines, at 4921.929 \ang and 5015.675 \ang, but also with Fe lI lines at 4923.92 \ang and 5018.44 \ang. The third line is consistent with the Fe\,{\sc ii} line at 5169.03 \ang, which favors an interpretation of the first three lines as an Fe\,{\sc ii} triplet. The fourth weak line can be identified with Mg\,{\sc i} at 5183.604 \ang. The insert shows a zoom-in of the region of the Fe\,{\sc ii} triplet and the Mg\,{\sc i} line.}
\label{aver}
\end{figure*}

\begin{table*}
\begin{center}
\begin{tabular}{rlllll}
a)  & & & &  \\
 & $\lambda _0$(\ang) & FWHM ($\rmn{km\,s^{-1}}$)  & Normalisation($\rmn{\AA}^2$) & Offset ($\rmn{km\,s^{-1}}$) \\
\hline
& H$\gamma$(4340.465) & 267$\pm$10 & 21.9$\pm0.71$ & 33.1$\pm8.08$ \\
& H$\beta$(4861.327) & 282$\pm$7 & 29.9$\pm0.7$ & 42.5$\pm6.9$ \\
& Fe\,{\sc ii}(4923.92) & 246$\pm$28 & 5.6$\pm0.55$ & 3.5$\pm12$ \\
& Fe\,{\sc ii}(5018.44) & 366$\pm$50 & 5.7$\pm0.61$ & -50$\pm21$ \\
& Fe\,{\sc ii}(5169.03)$^1$ & 410$\pm$36 & 8.4$\pm0.66$ & 20$\pm15$ \\
& Fe\,{\sc ii}(5169.03)$^2$ & 311$\pm$41 & 6.9$\pm0.77$ & -15$\pm11$ \\
\hline
%&  &  &  & \\
b) & & & & \\ 
&       & FWHM ($\rmn{km\,s^{-1}}$) & $v\sin i$$_\mathrm{em}$($\rmn{km\,s^{-1}}$)& Offset($\rmn{km\,s^{-1}}$)\\
\hline
      &Fe\,{\sc ii} group & 304$\pm$20 & 255$\pm$22 & -15$\pm11$ \\
 \hline                        
  \end{tabular}
  \end{center}
\caption{ \label{fittab} a) Results from the fitting of individual emission lines in the average spectrum of \src (see Figure \ref{aver}) with a Gaussian function. The wavelength of each line in the rest frame $\lambda _0$ (column 1) is frozen in the fit. The last column in the Table indicates the velocity offset of the fitted line centroid respect to $\lambda _0$. There is evidence for the presence of a narrow spike on the red wing of the Fe\,{\sc ii}(5169.03) line:  we fitted it without masking the spike ($^1$) and masking it  ($^2$) (see Section \ref{aversec} for a discussion). b) Results of the combined multi-Gaussian fit of the three Fe\,{\sc ii} lines, fitted together forcing the same offset  and FWHM in $\rmn{km\,s^{-1}}$. The spike affecting Fe\,{\sc ii}(5169.03) is masked. The $v\sin i$$_\mathrm{em}$ is measured by artificially broadening Arc lines with a Grey profile until reaching the FWHM measured for the \src lines. This accounts for the instrumental resolution profile. The line smearing due to orbital motion during the integration time is also taken into account. }  
\end{table*}

\begin{table*}
\begin{center}
\begin{tabular}{llllll}
 $\lambda _0$(\ang) & FWHM ($\rmn{km\,s^{-1}}$)  & Normalisation($\rmn{\AA}^2$) & Offset ($\rmn{km\,s^{-1}}$) &$v\sin i$$_\mathrm{em}$($\rmn{km\,s^{-1}}$) \\
\hline
Mg\,{\sc i}(5183.604) & 172$\pm$39 & 2.2$\pm0.4$ & 8$\pm16$ & 98$\pm$39 \\
 \hline                        
\end{tabular}
  \end{center}
\caption{ \label{Mgfit} 
Best parameters from the Gaussian fit of the faint emission line identifiable with Mg\,{\sc i} (5183.604 \ang) in the average spectrum (see Figure \ref{aver}). The last column reports the intrinsic line width $v\sin i$$_\mathrm{em}$, measured as for the Fe\,{\sc ii} lines (see caption Tab. \ref{fittab}). } 
\end{table*}

\subsection{Averaged spectrum}
\label{aversec}

In order to measure the FWHM of the emission lines, we averaged the spectra in the frame of the companion star. We assumed a circular orbit, shifting the lines by $v= -$K$_\mathrm{em}sin(2\pi\phi+\varphi)$, with $\varphi=0$.  We did not adopt a priori the K$_\mathrm{em}$ obtained from the radial velocity curves of the Balmer lines, since that could be affected by, e.g., asymmetry or variations in the line profile from the individual spectra (as for the H$\alpha$, \citealt{2009MNRAS.399.2055B}). Moreover, the Balmer lines could originate in a different area than the weaker Fe\,{\sc ii} ones and thus have a different K$_\mathrm{em}$. Instead, we proceeded as follows: 
\begin{itemize} \item[-] we assumed a range of possible values of K$_\mathrm{em}$ (200$<K_D<$400$~\rmn{km\,s^{-1}}$  in steps of 10 $\rmn{km\,s^{-1}}$). For each K$_\mathrm{em}$ value we applied the circular orbital shift and averaged  the spectra between phase 0.3 and 0.7 (the range where the weaker emission lines are clearly detected). 
\item[-] we measured the FWHM of the emission lines in the averaged spectra, fitting a Gaussian function to each line (see Section \ref{features}).  An orbital shift close to the real one will result in narrower lines in the averaged spectrum. 
\item[-] we found the K$_\mathrm{em}$ that results in the narrowest lines: for each line, we plotted the width measured from each average spectrum against the K$_\mathrm{em}$ of the corresponding orbital shift. The uncertainty on the width was large with respect to its variation among different values of K$_\mathrm{em}$. Nonetheless, each line displays a trend indicating a minimum width at K$_\mathrm{em} \sim 300~ \rmn{km\,s^{-1}}$. The result is consistent with that from the radial velocity curves in Section \ref{rvcsec}.
\end{itemize}
The spectrum in Figure \ref{aver} is the average of individual exposures comprised between phase 0.35 and 0.7 (where all the emission lines are visible) for K$_\mathrm{em}=300~\rmn{km\,s^{-1}}$. Together with H$\beta$ and H$\gamma$, the averaged spectrum highlights the weak Fe (or He) emission lines. 

\noindent No absorption lines appear in the averaged spectrum. As for the individual spectra, a cross-correlation of the average spectrum with the standard template spectra does not give a match. Note that we have also constructed an averaged spectrum in the phase interval 0.8 to 0.2 and  0.9 to 0.1, but we did not detect absorption lines from the non-irradiated face of the companion star. 

\noindent Table \ref{fittab} presents the parameters of the best-fitting Gaussian to each line in the averaged spectrum.  Each line was fitted for FWHM, normalisation and for the offset $\lambda-\lambda_0$ of the line centroid $\lambda$ with respect to the rest-frame wavelength $\lambda _0$. 
As we corrected the spectra for the orbital motion only, the lines in the average spectrum are still shifted with respect to their rest-frame wavelength by the systemic radial velocity $\gamma$. The wavelength offset  of the H$\beta$ and the H$\gamma$ lines is consistent on a 2$\sigma$ level with the values of $\gamma$ derived from the radial velocity curves (see Section \ref{rvcsec}). The weighted average of the messures of $\gamma$ from the radial velocity curves and  from the offsets of the Balmer lines is $\gamma=$43.8$\pm$3.6$~\rmn{km\,s^{-1}}$. The offsets of the Fe\,{\sc ii}  lines are not consistent with this value, although they agree with each other on a  3$\sigma$ level.  In order to understand this difference, we have tested the dependence of the measured line offsets in the averaged spectrum on the choice of K$_\mathrm{em}$ and  $\varphi$ in the orbital motion correction. If the Fe\,{\sc ii} lines originate in different regions of the companion star, in fact, the two sets of lines will be associated with a different K$_\mathrm{em}$ and  $\varphi$. We found that the offset changes slowly with K$_\mathrm{em}$, but  is sensitive to variations in  $\varphi$. 
%This is easily understood as a constant phase difference will translate in a constant offset.
The offset of the Fe\,{\sc ii} lines in the average spectrum is  consistent  with the measure of $\gamma$ from the Balmer lines if the orbital motion of their source region is shifted in phase by $-$0.03$\lesssim \varphi \lesssim-$0.05. 

\noindent As we previously pointed out, the FWHM of the lines  (both H$\gamma$, H$\beta$ and the Fe\,{\sc ii} group) is not affected by changes in K$_\mathrm{em}$ of a few tens of $\rmn{km\,s^{-1}}$ around $\sim$300 $\rmn{km\,s^{-1}}$. Unlike the offset, it is also not sensitive to changes of a few percent in $\varphi$. In other words, our measure of the FWHM is not affected by a possible small displacement of the Fe\,{\sc ii} source region with respect to the source region of the Balmer lines. All the lines in the average spectrum have the same FWHM at a 2$\sigma$ level, with the exception of the reddest Fe line, Fe (5169.03 \ang), which is significantly broader than Fe (4923.92 \ang) at the $>$3$\sigma$ level. Careful inspection shows a narrow faint emission line on the  red end of the 5169.03 \ang line (Figure \ref{aver}). The presence of this second peak artificially broadens the Gaussian function used to fit the Fe (5169.03 \ang) line, as the fitting routine tries to account for both peaks with the same Gaussian. If this narrow peak is masked or fitted with an additional line, the width of  Fe (5169.03 \ang) becomes consistent within 1$\sigma$ with  the other lines. It must be noted that a second Gaussian introduced to fit the faint emission near the Fe (5169.03 \ang) line helps the fit  but is not significant itself (the normalisation is approximately equal to the uncertainty on it). 
After we have verified that the Gaussian fit to the Fe  {\sc ii} lines separately provides FWHMs and offsets in agreement with each other,  we fitted the triplet together in order to reduce the uncertainty on the parameters.  The three Fe lines were forced to have the same FWHM in $\rmn{km\,s^{-1}}$ and the same velocity offset. The faint emission on the reddest Fe line was masked to avoid artificial broadening.  We obtained a FWHM=$304\pm22~\rmn{km\,s^{-1}}$ (reduced $\chi^2=0.99$, 994 d.o.f. ).  

\subsection{Measuring the intrinsic broadening of the emission lines}
\label{broad}
\noindent The observed FWHM of the emission lines is determined by the intrinsic line width broadened by the instrumental resolution profile and by line smearing produced by the motion of the  companion during the integration time of one spectrum. Intrinsic saturation effects can also contribute to the width of the emission lines, significantly affecting strong  lines like the Balmer ones  \citep{1989agna.book.....O}. On the other hand,  if the weak Fe\,{\sc ii} lines originate on the companion star, their intrinsic width is expected to be dominated by the rotational broadening  $v\sin i$$_\mathrm{em}$ \citep{1992oasp.book.....G}. In order to measure the latter from the Fe\,{\sc ii} triplet, we have artificially smeared and broadened the arc spectra with different values of  $v\sin i$$_\mathrm{em}$. The $v\sin i$$_\mathrm{em}$ resulting in smeared arc lines as broad as the FWHM of the Fe\,{\sc ii} lines is a measure of the rotational broadening. We proceeded as follows:  the phase of each \src spectrum  participating in the average in Figure \ref{aver} was ascribed to one arc spectrum. We smeared the arc spectra 
%with the shift in velocity suffered by the corresponding (same phase) \src spectrum, given 
by $2 \pi T K_\mathrm{em} \cos(2\pi\phi)/P$ where T is the duration of one exposure (900 s) and P the orbital period. The smeared arcs were then averaged,  each one with the same weight as that of  the corresponding (same phase) \src spectrum in the average of Figure \ref{aver}.  In this way we have simulated the effect of the smearing due to the orbital motion on the average spectrum. We then broadened the smeared average arc spectrum with a Grey profile for different values of  $v\sin i$$_\mathrm{em}$, measuring each time the resulting FWHM in $\rmn {km\,s^{-1}}$. We find that the measured FWHM of the Fe\,{\sc ii} lines corresponds to $v\sin i$$_\mathrm{em}$= 255$\pm$ 20 $\rmn{km\,s^{-1}}$.

\subsection{Evidence for Mg{\sc i} in emission}
\label{mg}
The average spectrum displays a weak emission line consistent with Mg\,{\sc i} at  5183.604 \ang. The line was not detectable in the single or trailed spectra, but a fit with a single Gaussian function indicate that the line is significant on a 5$\sigma$ level in the average spectrum. Two other Mg\,{\sc i} lines, at 5167.321 \ang and 5172.684 \ang, belong to the same multiplet with Mg\,{\sc i} (5183.604 \ang). The position of Mg {\i} (5172.684 \ang) is consistent with that of the spike on the red wing of Fe\,{\sc ii} (5169.03 \ang) (see Fig. \ref{aver}), although the detection of the latter is not statistically significant when the prominent Fe\,{\sc ii} line is included in the fit.  Mg\,{\sc i} (5167.321 \ang)  could be present but it can not be resolved from Fe\,{\sc ii} (5169.03 \ang). The FWHM of the Mg\,{\sc i} (5183.604 \ang) line is 172$\pm$39 $\rmn {km\,s^{-1}}$, which corresponds to an intrinsic width of  $v\sin i$$_\mathrm{em}$= 98$\pm$ 39 $\rmn{km\,s^{-1}}$ (measured as we did for the Fe\,{\sc ii} triplet in Section \ref{broad}).  This value is smaller than what we obtained from the Fe\,{\sc ii} triplet.

\subsection{Doppler tomography}
\label{Doppler}

\begin{figure}
\centering
\begin{tabular}{c}
H$\beta$ \\
\includegraphics[width=6.0cm, angle=0]{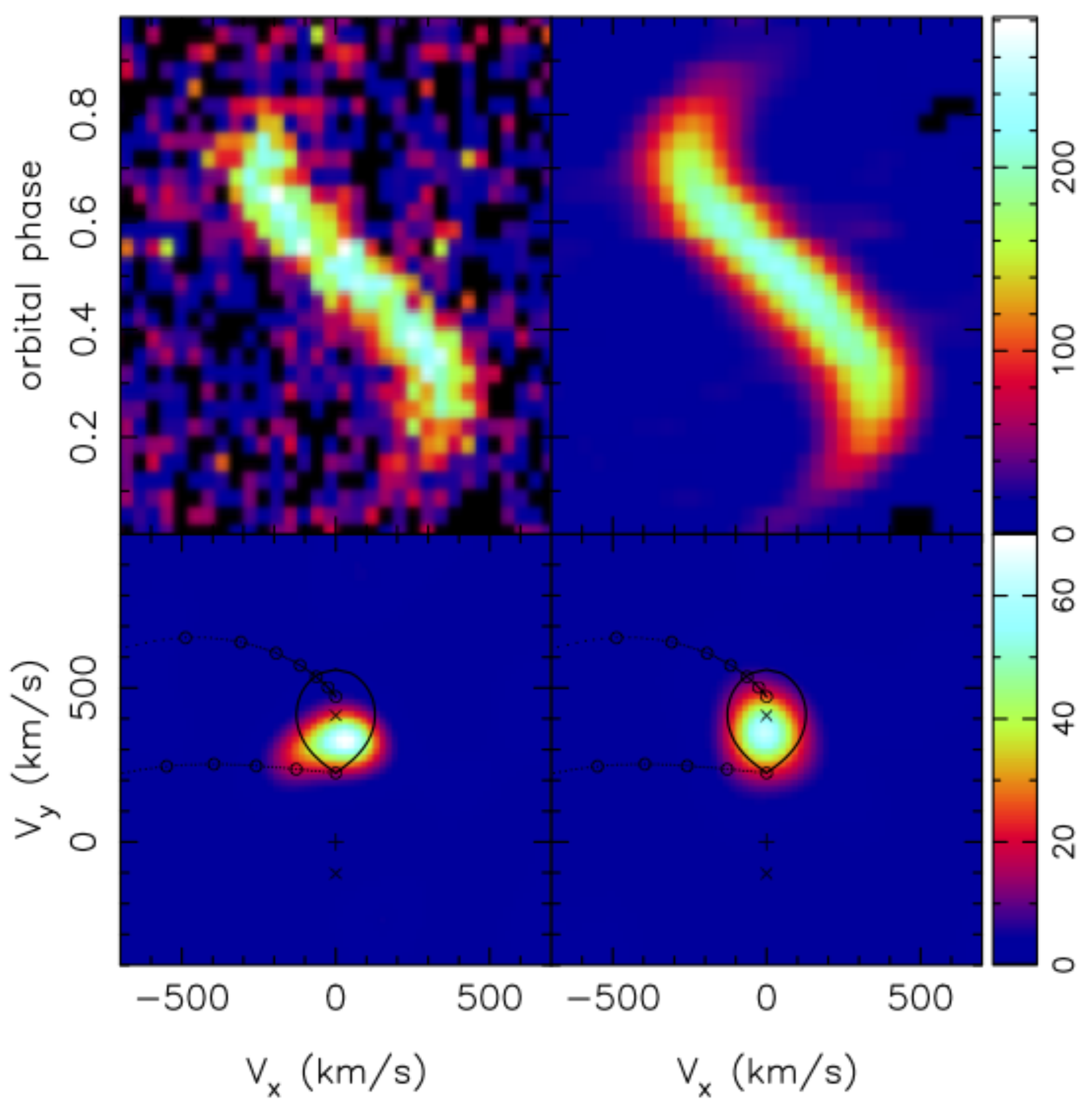}  \\
Fe\,{\sc ii}\,(4923.92\,\ang)  \\
\includegraphics[width=6.0cm, angle=0]{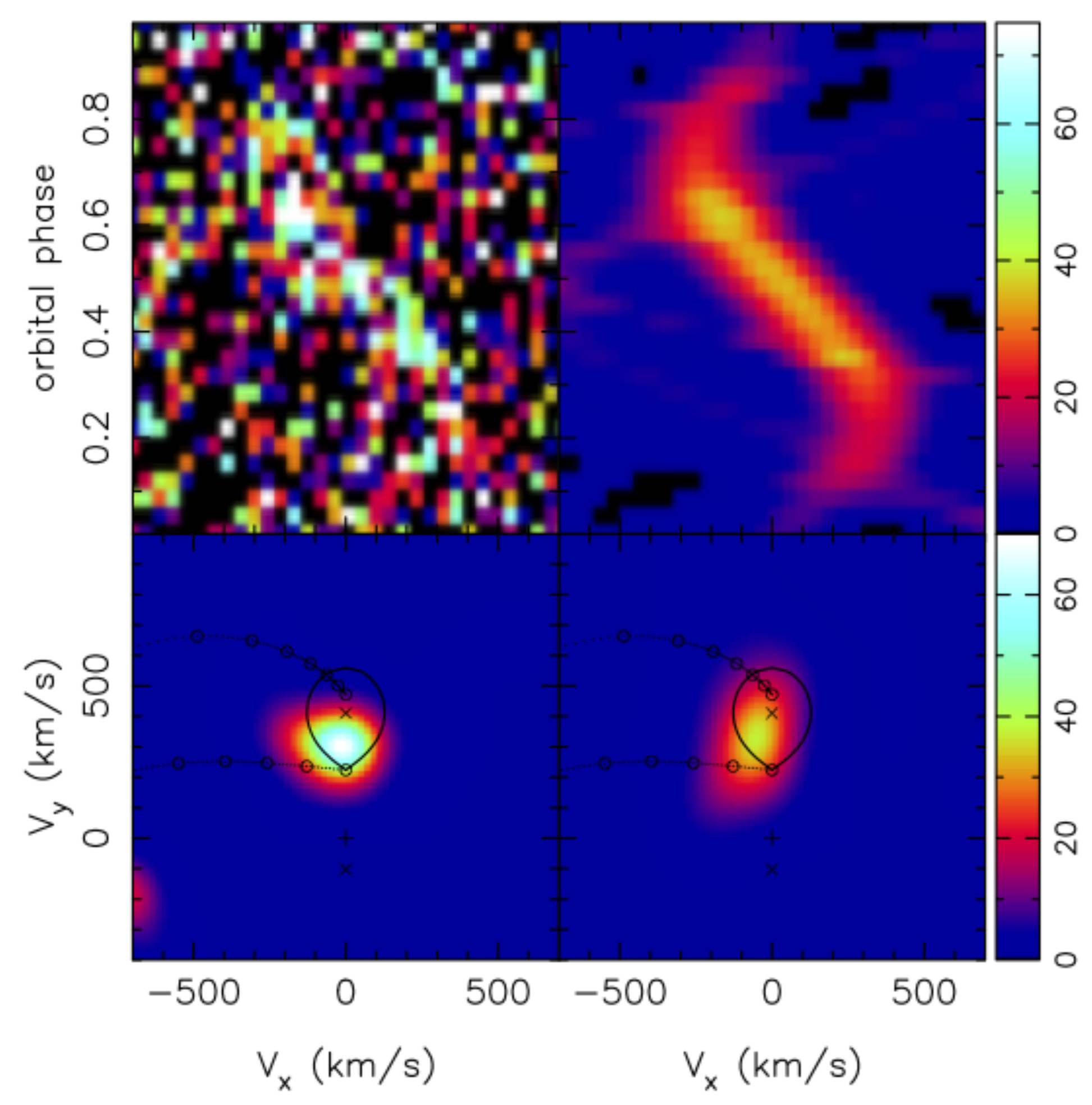} \\
Fe\,{\sc ii}\,(5169.03\,\ang) \\
\includegraphics[width=6.0cm, angle=0]{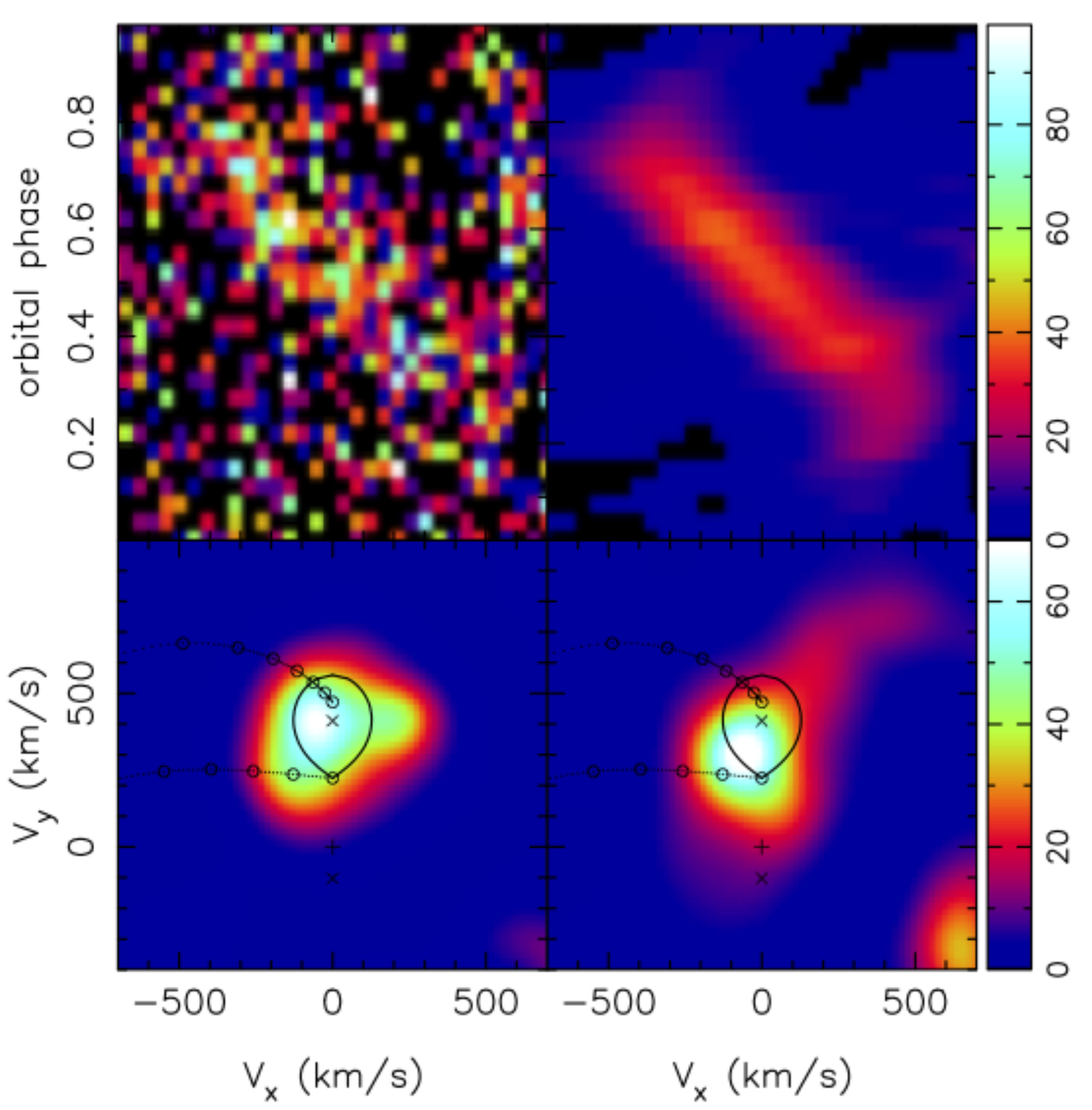} \\
\end{tabular}
\caption{Tomographic reconstructions for the H$\beta$ and for the Fe\,{\sc ii} lines at 4923.92 \ang and 5169.03 \ang. The top-left panel in each case shows the observed line profile as a function of the orbital phase while the top-right panel shows the reconstructed data from the converged maximum entropy solution. The bottom panels show the Doppler maps resolving the emission of each line in a V$_X-$V$_Y$ plane. The constant and the phase-dependent contribution to the flux are shown in the left and right quadrant respectively (see \citealt{2003MNRAS.344..448S}). The color scale indicates the fractional amplitude of the variable map in per cent. A Roche lobe model and ballistic stream trajectory is plotted for an assumed system mass ratio of q=0.25 and radial velocity semi-amplitude of the companion star K$_2=$410 $\rmn{km\,s^{-1}}$.  }
\label{dmaps}
\end{figure}

%The emission component is a very pure modulated s-wave, with almost 100% modulation amplitude. Good fits are achieved since its a very clean signal, and clearly must be from near the front of the lobe to achieve such large modulation. Also, the phase is pretty much all in the cosine, i.e. peaking at 0.5, as expected for a centered emission spot on the lobe.The peak amplitudes are ~330 km\,s^{-1}, a bit lower than what H-alpha showed in the DDT data. There is no evidence at all for an accretion disc in emission, as we saw faintly in the DDT data. 

%Here are some Doppler maps with Roche lobes (k2=410, q=0.25) and in color that might work for presentation.  Hbeta looks good though closer to L1 than Halpha during the DDT data. He4921 is indeed most likelt Fe4923 instead, giving a similar map, but data is noisy. The 5169 feature may be consistent with Fe\,{\sc ii}, but seems broader and causes a rather extended blob in the map. So beware when measuring vsini from it. Also all spots are shifted slighlty to the left, so maybe I did mess up my ephemeris after all...

We employed emission line Doppler tomography to map the observed emission features. Here we used the modulation Doppler tomography method of \citet{2003MNRAS.344..448S}, using the same code that was employed in the \citet{2009MNRAS.399.2055B}
study. The observed spectra were first phase binned using 30 orbital bins. The underlying continuum was subtracted using a polynomial fit to line-free regions. In Figure \ref{dmaps} we show the resulting tomograms for the H$\beta$ line and the two Fe\,{\sc ii} lines at 4923.92 and 5169.03 \ang. The H$\gamma$ line reconstructions were effectively identical to H$\beta$ and we compare the
two Fe\,{\sc ii} lines given their apparent difference in intrinsic width. All emission lines reconstructions are consistent with line emission from a region near the mass donor star with no evidence for any extended emission from a residual accretion disc or stream. We find that the line flux is
strongly modulated, as expected for an origin on the irradiated face of the companion, and in phase with the observed continuum modulation. Good fits are achieved close to a reduced $\chi^2$ of 1.

\noindent The Doppler maps confirm the key attributes we derived in the previous sections using Gaussian fitting. We find that the radial velocity amplitude
of the emission ranges from 300-350 $\rmn{km\,s^{-1}}$, consistent with the $K_\mathrm{em}$ values from Section \ref{aversec} as well as past estimates of $K_2$.  We show in Figure \ref{dmaps} the expected location of the donor Roche lobe and gas stream using a mass ratio of q=0.25 (consistent with the NS mass range favoured by \citealt{2006Natur.441.1115O} and the work of \citealt{MunozDarias:2009p73}, see also  \citealt{2009MNRAS.399.2055B}) and $K_2$=410 $\rmn{km\,s^{-1}}$. The latter is the lower limit to $K_2$ obtained by \citet{2009MNRAS.399.2055B} which is a more stringent constraint  compared to what we derived in the previous sections (see the discussion in section \ref{discussion}). Again, the line at 5169.03 \ang  appears somewhat anomalous with a significantly more extended emission distribution, reflecting its observed width. Furthermore, the center of the emission is shifted slightly off the Roche lobe, consistent with the
apparent phase shift mentioned previously.

\noindent In summary, our Doppler tomograms confirm that the narrow emission lines are consistent with an origin on the irradiated face of the mass donor
star, with the exception of the emission feature at 5169.03 \ang which is both extended and shifted with respect to the other lines. This is in agreement  with the broadening effect observed in the Gaussian fitting of 5169.03 \ang (Section \ref{aversec}). 
We will discuss this further in Section \ref{discussion}.

\section{Summary of the results}
\label{summary}

The results presented in the previous sections can be summarised as follows: 

\begin{itemize}

\item[-] The g'-band light curve of the source spans a range of ~1 magnitude. The profile is variable, but the errors on the individual magnitudes are large: significant variability  at a 3$\sigma$ level is detected within the phase interval of 0.007 starting from phase 0.349, with an amplitude of  0.35$\pm0.08$ magnitudes (Figure \ref{trail} top panel ).  The morphology of the light curve in the g'-band is similar to that of the r'-band light curve reported by \citet{2009MNRAS.399.2055B}, albeit the latter is not variable. 

\item[-] The spectra show emission lines from H$\beta$ and H$\gamma$ and three weaker lines consistent with Fe\,{\sc ii} (4923.92 \ang) or He\,{\sc i} (4921.929 \ang), Fe\,{\sc ii} (5018.44 \ang) or He\,{\sc i} (5015.675 \ang) and Fe\,{\sc ii} (5169.03 \ang). Since there is no He\,{\sc i} line at the position of the last weak emission feature, we favour an Fe\,{\sc ii} group interpretation.  A weak emission line consistent with Mg\,{\sc i} (5183.604) is also detected at a 5$\sigma$ level in the average spectrum. 

\item[-] The orbital motion displayed by the emission lines is consistent with them originating close to the surface of the companion star facing the NS. The phasing of the continuum variation is in sync with the strong orbital modulation of the emission line strength, as expected if both are mainly produced by the irradiated companion.  Evidence of an irradiated companion star was previously found, both during the X-ray outburst phase (\citet{Pearson:2006p241}, \citealt{MunozDarias:2009p73}) and at the beginning of the quiescent period \citep{2009MNRAS.399.2055B}.

\item[-]  The broad disc component at the base of the emission lines that was detected in the spectra from \citet{2009MNRAS.399.2055B} is not detected here. There is no  evidence for an accretion disc either from the line profiles or the Doppler tomography.

\item[-]  The fitting of the radial velocity curves of the H$\gamma$ and H$\beta$ lines  provides 1 $\sigma$ consistent measures of the radial velocity amplitude of the emission lines K$_\mathrm{em}$, whose weighted average is 308.5$\pm$3.9 $\rmn{km\,s^{-1}}$. The minimum width of the emission lines in the average spectrum is obtained with an orbital-motion correction with K$_\mathrm{em}\sim$300 $\rmn{km\,s^{-1}}$, in agreement with the radial velocity curves estimate.  This measurement provides a lower limit on the radial velocity semi-amplitude of the companion star $K_2$, which is fully consistent with the results of \citet{MunozDarias:2009p73}. A more stringent lower limit on $K_2$ is reported by \citet{2009MNRAS.399.2055B} (see discussion in section \ref{discussion}).

\item[-]  The systemic radial velocity $\gamma$ was measured from the radial velocity curve traced by the H$\gamma$ and H$\beta$ lines and from the wavelength offset of the same lines in the orbital-motion subtracted averaged spectrum. The resulting values of $\gamma$ are in agreement  within 2$\sigma$, their weighted average being 43.8$\pm$3.6 $\rmn{km\,s^{-1}}$. The offset of the Fe\,{\sc ii} triplet in the average spectrum, instead, is inconsistent with this value of  $\gamma$. A phase shift of -0.03/-0.05 of the source region of the Fe\,{\sc ii} lines with respect to the binary axis can account for this discrepancy (see Section \ref{discussion}). 
%This can be explained with a displacement of the region where the Fe\,{\sc ii} triplet is produced with respect to the Balmer lines source region. In particular, the  Fe\,{\sc ii} source region should be slightly off with respect to the binary axis, with a phase shift of -0.03/-0.05(see Section \ref{discussion}). 
The Doppler maps also show that the emission of the Fe\,{\sc ii} lines is shifted, particularly for the one at 5169.03 \ang.
Our measure of the systemic radial velocity is in agreement with the work of \citet{MunozDarias:2009p73}, but is more than 5$\sigma$ apart from the value obtained by \citet{2009MNRAS.399.2055B}, $\gamma=$78.6$\pm$3.9 $\rmn{km\,s^{-1}}$.  Even taking into account  that the value of $\gamma$ is affected by the accuracy of the wavelength calibration (0.023 \ang in \citealt{2009MNRAS.399.2055B}  and 0.1 \ang for our data) the two measurements differ by more than 4$\sigma$. Similar mismatches using VLT data are not unprecedented (see  \citealt{Oro11}).

% We do not foresee a clear reason for this discrepancy.
 
\item[-] With the exception of the weak Mg\,{\sc I}, the emission lines are very broad. The FWHM of the Fe\,{\sc ii} lines indicates a $v\sin i$$_\mathrm{em}$= 255$\pm$ 22 $\rmn{km\,s^{-1}}$, if the broadening results from the rotation of the companion star alone. The Mg\,{\sc I} (5183.604 \ang) line instead provides a much lower $v\sin i$$_\mathrm{em}$ of 98$\pm$ 39 $\rmn{km\,s^{-1}}$.

\end{itemize}

\section{Discussion}
\label{discussion}

Using medium resolution VLT FORS2 spectra of \src, ranging from 4222 to 5701 \ang, we found emission lines  indicating  that the companion star is still heated by irradiation, as was observed soon after the onset of the quiescent phase  \citep{2009MNRAS.399.2055B}.  We detected H$\gamma$, H$\beta$ and a set of Fe\,{\sc ii} lines, which is unusual but not  unprecedented in LMXBs \citep{1994MNRAS.266..137M}. We also detect a faint Mg\,{\sc i} line. The source of the irradiation is most likely the X-rays associated with the cooling compact object \citep{2009MNRAS.396L..26D}. The slow rate of the NS cooling that has been observed  in \src with ${\it XMM}$, ${\it Chandra}$ and ${\it Swift}$  (\citealt{2010arXiv1007.0247D}, \citealt{ 2011A&A...528A.150D}) is in agreement with the fact that the companion star is still significantly irradiated, after more than one year of quiescence. 

\noindent The g'-band light curve of \src shows evidence for variability between different orbits, in the phase interval of 0.007 starting from phase 0.349. The variability amplitude is 0.35$\pm0.08$ magnitudes. This implies a change in the flux heating the companion star, which could potentially be due to a change in a pulsar wind (see below) or to variability in the X-ray radiation, although the latter is hard to explain given the small variations observed in the X-rays (\citealt{2010arXiv1007.0247D}, \citealt{ 2011A&A...528A.150D}). Furthermore, theoretically one does not expect the X-ray flux of the glowing cooling NS to vary on short timescales \citep{2002ApJ...574..920B}.
A comparison with the R-band light curve presented in \citet{2009MNRAS.399.2055B} shows that the  morphology of the light curves is similar, albeit the g'-band one is variable. 

\noindent  Although we can not rule out a disc contribution to the optical continuum, the non-detection of disc line emission is unusual in comparison with typical observations of quiescent LMXB (e.g. \citealt{1994MNRAS.266..137M}, \citealt{ 2002MNRAS.334..233T}). The disc-instability model that can explain the short outburst of typical X-ray transients (\citealt{1971AcA....21...15S}, \citealt{2001A&A...373..251D}) predicts that the disc is present even during quiescence and that an outburst is triggered by a sudden rise of viscosity. The lack of disc lines in \src suggests instead that the mass transfer has dramatically dropped and might have even stopped since the observations performed by \citet{2009MNRAS.399.2055B} at the beginning of the quiescent phase. On the other hand, the disc-instability model might not apply to the case of \src as it can not easily explain the 20-year long outburst that the source underwent. The trigger of tens of years long outbursts is not well understood and might have to do with variations in the envelope of the companion star, rather than with instabilities of an accretion disc \citep{2006ARA&A..44...49R}.

\noindent In an attempt to obtain constraints on the NS mass from the emission lines,  we determined the rotational velocity obtained from the FWHM of the Fe\,{\sc ii} triplet observed in the spectra. If those lines are produced on the companion star surface, their width is expected to be dominated by the effect of rotational broadening  $v\sin i_\mathrm{em}$ \citep{1992oasp.book.....G}. As also emerged from the tomographic reconstructions, the line emission does not arise from the full inner hemisphere of the companion star, but concentrates close to L1 point. For this reason the  $v\sin i$$_\mathrm{em}$ obtained from the Fe\,{\sc ii} lines would represent a lower limit to the projected spin velocity of the star $v\sin i$.  Due to the shift between the center of light of the Fe emission lines and the center of mass of the companion, the radial velocity semi-amplitude of the emission lines K$_\mathrm{em}$ also provides only a lower limit to K$_D$. The combination of the lower limits on K$_D$ and $v\sin i$ provides in turn a lower limit on the NS mass. From the radial velocity curves and from our average spectrum we measured K$_2 \gtrsim$ 308.5$\pm$3.9 $\rmn{km\,s^{-1}}$ (see Sections \ref{rvcsec} and  \ref{aversec}). From the Doppler maps (which are not influenced by asymmetries in the line profiles, see \citealt{2009MNRAS.399.2055B}) we obtain a higher limit, K$_2 \gtrsim$355 $\rmn{km\,s^{-1}}$. An even more stringent limit is provided in \citet{2009MNRAS.399.2055B}, K$_2\gtrsim405~\rmn{km\,s^{-1}}$. This was obtained from Doppler tomography of the H$\alpha$ line, which is expected to form  closer to the center of mass of the companion star with respect to higher ionization potential lines such as the H$\beta$ and H$\gamma$, providing a radial velocity closer to K$_D$ (e.g. \citealt{1999A&A...341..491H}, \citealt{2006MNRAS.369..805U}).

\noindent The constraints on the mass of the NS in \src that can be obtained combining the results of \citet{2009MNRAS.399.2055B} with our measure of $v\sin i$$_\mathrm{em}$ show that, if the width of the Fe lines in our spectra is indeed setting a lower limit to $v\sin i$, then the mass of the compact object in \src should be $\gtrsim$3.5 \msun. This is more than the value of $\sim$3 \msun which is considered to be an upper bound to the mass of a NS. A  BH interpretation would be in conflict with the detection of thermonuclear X-ray bursts from the source, which establish that the compact object is a NS (\citealt{gohapa1986}). Given the fact that a bursting BH or a 3.5\msun NS would be against theories about those objects, we have investigated possible scenarios to solve this incongruity.

\noindent Our mass constraint is based on the assumption that the width of the Fe lines provides a measure of the projected rotational velocity of the companion star $v\sin i$, but that might not be the case. An asymmetric irradiation of the companion due, e.g., to the shadowing of a residual accretion disc, could induce currents in the atmosphere trying to re-distribute the heat (\citealt{1982MNRAS.200..907K}, \citealt{2002ApJ...575..384W}). The velocity of the current would add to the width of the emission lines. Nevertheless, this velocity has to be of the order of the total line broadening in order to give a significant contribution, i.e. a few hundreds $\rmn{km\,s^{-1}}$.  This is more than the sound speed on the surface of a late type star \citep{2002apa..book.....F}. 

\noindent Another possibility is that the Balmer lines, originating closer to the companion star surface, are broadened by saturation effects \citep{1989agna.book.....O}, while the Fe\,{\sc ii} triplet, slightly phase shifted, is produced in a stream of residual accreting matter whose velocity causes the observed FWHM. This interpretation requires a bit of fine tuning of two broadening effects and it does not fit well with the lack of any evidence for an accretion disc.
 
\noindent The large width of the Fe lines could also be due to a blend with unresolved lines of other atomic species. This is suggested by the detection of a faint (5$\sigma$ significance) line consistent with Mg\,{\sc i} at  5183.604 \ang. As mentioned in Section \ref{mg}, the latter belongs to a Mg\,{\sc i} multiplet including lines at 5167.321 \ang and 5172.684 \ang, both very close to the position of the reddest detected Fe\,{\sc ii} line. Mg\,{\sc i} (5172.684 \ang) might be identified with the peak next to Fe\,{\sc ii} (5169.03 \ang) that we have masked when measuring the FWHM of the Fe\,{\sc ii} line. The peak is not significant in the average spectrum, but the anomalous extension and shift of Fe\,{\sc ii} (5169.03 \ang) in the Doppler maps suggest that this narrow feature might be real and present in all the spectra, altering the look of the line in the Tomographic reconstruction. On the other hand,  Mg\,{\sc i} (5167.321 \ang) could be present, but it is not resolved from Fe\,{\sc ii} (5169.03 \ang). While a blend with Mg\,{\sc i} can explain the width of  Fe\,{\sc ii} (5169.03 \ang), a blend with He\,{\sc i} can broaden the other two lines of the Fe\,{\sc ii} triplet. As in the "stream" scenario, the FWHM of H$\beta$ and H$\gamma$ would instead be determined by saturation effects. The downside of this sketch is that it requires some fine tuning of the width and intensity of the lines of different species in order to produce consistent FWHM and offsets among the three Fe\,{\sc ii} lines.  
%The saturation effect on the Balmer lines should be such to produce similar widths as well. Moreover, 
Other lines from Mg\,{\sc i} and He\,{\sc i} (at 4387.928 and 4471.68 \ang) that are within the wavelength coverage of our observation are not detected, so we have to meet the further requirement that only the lines that overlap with Fe {\sc II} are strong enough to be observed.  Although the  Mg\,{\sc i} (5183.604 \ang) detection supports this scenario, we can not confirm it or rule it out with the resolution and signal-to-noise of our observations. 

For this reason, we have considered another possible situation, where  the lines are produced in a wind of matter evaporating from the companion star, e.g. due to the action of a pulsar wind and/or the X-ray heating.  The emission lines, originating in the wind of evaporated material, are broadened due to the wind velocity. The Balmer lines, produced closer to the companion star surface, display the orbital motion of the irradiated secondary, while the Fe\,{\sc ii}, rising from further out in the wind, display a slight phase delay in the orbital motion as the evaporated material starts to trail the companion star as it moves outwards due to angular momentum conservation.  The evaporation of material could also explain the lack of evidence for emission lines from an accretion disk, as it would quench the accretion. Moreover, the presence of a pulsar wind in addition to the X-ray heating might be responsible for  the observed variability in the g'-band light curve, which is possible but rather difficult to produce in terms of variability of the X-ray flux. At last, this ``Black-widow like''  scenario is consistent with the extended distribution of the line emission in the Doppler maps. The weak point of this outline is that it may be difficult to produce the Mg\,{\sc i} (5183.604 \ang), which is narrower than the other lines. The weakness of the line did not allow the production of Doppler maps to verify if the source region of this feature is narrow or extended and if it is the same as the Balmer or Fe {\sc ii} lines. 
Moreover, it has to be noticed that the pulsar in quiescence has not been found in the radio waveband (private communication M. Burgay).  

\section{conclusions}

We performed phase-resolved medium-resolution optical spectroscopy with VLT/FORS2 on the eclipsing LMXB \src , one year and two months after it returned to quiescence.  We found evidence for variability in the g'-band light curve and we found that the spectra only present emission lines coming from the region close to the companion star, with no line contribution from a disc. This is unusual compared with typical X-ray transients and could be related to the mechanism that triggers long duration outbursts such as those displayed by the source, which is not yet understood. 

%It could also be that accretion is quenched by a pulsar wind evaporating the envelope of the companion star at the Lagrangian point.

\noindent The g'-band variability can be explained by a varying X-ray flux or by a variable pulsar wind.  X-ray observations show little variations in the flux, as expected from a cooling NS. 

%Again, an additional source of heating such as a pulsar wind could account for the observed variability. 

\noindent The H$\beta$, H$\gamma$ and Fe\,{\sc ii} emission lines in the average spectrum are remarkably broad. Their FWHM is not likely to be dominated by rotational broadening alone as this would lead to a compact object mass inconsistent with a NS.  The width of the lines can be explained with bulk gas motion, e.g. the lines could be produced in a wind of material evaporated from the companion star envelope due to X-ray heating, possibly in combination with a pulsar wind. It is also possible that some of the lines are blends of unresolved features. Despite extensive efforts to search for spectral features using multiple methods, we detect no absorption features from the mass donor star.
Deeper observations with better resolution might allow to resolve blends and detect more emission lines that could provide a reliable lower limit to the companion star rotational broadening and, in turn, to the NS mass. Moreover, there is still the chance to detect absorption lines from the non-irradiated face of the companion star,  which would finally provide a mass estimate for this elusive NS.

\section*{Acknowledgments} \noindent PGJ acknowledges support from a 
VIDI grant from the Netherlands Organisation for Scientific
Research. Tom Marsh is thanked for developing
and sharing his packages {\sc Pamela} and {\sc Molly}.

\bibliographystyle{mn} \bibliography{EXO.bib}

\end{document}